\documentclass[useAMS,usenatbib]{mn2e}

\title[Neutron Star Magnetism]{Type I and Two-Gap Superconductivity in
Neutron Star Magnetism}
\author[P. B. Jones]{P. B. Jones\thanks{E-mail:
p.jones1@physics.ox.ac.uk}\\
Department of Physics, University of Oxford, Denys Wilkinson Building,
Keble Road, Oxford OX1 3RH\\}

\begin{document}

\date{}

\pagerange{\pageref{firstpage}--\pageref{lastpage}} \pubyear{}

\maketitle

\label{firstpage}

\begin{abstract}
Neutron-star inner cores with several charged baryonic components
are likely to be analogues of the two-gap superconductor which is
of current interest in condensed-matter physics.  Consequently,
type I superconductivity is less probable than type II but may
nevertheless be present in some intervals of matter density.
The intermediate-state structure formed at finite magnetic
flux densities after the superconducting transitions is subject
to buoyancy, frictional and neutron-vortex interaction forces.
These are estimated and it is shown that the most important
frictional force is that produced by the stable stratification of
neutron-star matter, the irreversible process being diffusion in
the normal, finite magnetic-flux density, parts of the structure.
The length-scale of the structure, in directions perpendicular to
the local magnetic field is of crucial importance.  For small
scales, the flux comoves with the neutron vortices, as do the
proton vortices of a type II superconductor.  But for much larger
length-scales, flux movement tends to that expected for normal
charged Fermi systems.
\end{abstract}


\section{Introduction}

The evolution of neutron star magnetic fields has been a topic of
considerable interest since the early papers of Pacini (1967) and
Gold (1968), the discovery of pulsars by Hewish et al (1968), and
the paper of Baym, Pethick \& Pines (1969) on proton type II
superconductivity. There has been much uncertainty in the magnetic
flux transport properties of those parts of the interior where the
density does not exceed the nuclear density $\rho_{0} = 2.5 \times
10^{14}$ g cm$^{-3}$.  For a typical neutron star of mass 
$1.4 M_{\odot}$, the central density predicted by many equations of
state is $\rho\sim 2-4\rho_{0}$, indicating the presence of hyperons
and of even greater uncertainties in magnetic flux transport
properties in the inner core.  It is evident that empirical deduction
of these properties is not feasible and that {\it a priori} theoretical
input is required.

The general assumption has been that the protons form a $^{1}S_{0}$
type II superconductor.  As functions of matter density, calculated
proton energy gaps are typically $\Delta_{p}\sim 0.5$ MeV
at $\rho\simeq\rho_{0}$ but decrease rapidly at higher densities
(see Pethick \& Ravenhall 1995; Heiselberg \& Hjorth-Jensen 2000). Thus
it seems likely that the condition for type II superconductivity,
$\kappa=\lambda/\xi > 1/\sqrt{2}$, where $\lambda$ is the penetration
depth and $\xi$ the coherence length, will be satisfied at
$\rho\simeq\rho_{0}$,
that is, in regions of the outer core where matter is composed of
protons and $^{1}S_{0}$ or $^{3}P_{2}$ superfluid neutrons,
charge-neutralized and in weak-interaction equilibrium with electrons
and negative $\mu$-mesons.  Calculations of $\Delta_{p}$ are even less
reliable at higher densities, $\rho\sim 2\rho_{0}$, so that the
possibility of type I superconductivity in the inner core cannot be
excluded.  There has been some work on $\Sigma^{-}$-hyperon pairing
(see Vida\~{n}a \& Tol\'{o}s 2004) which indicates a large
$^{1}S_{0}$ gap and the possibility that the inner-core superconductor
has two or more components.  But much uncertainty remains concerning
all baryonic gap calculations, and some authors have questioned the
existence of type II superconductivity at any core density
(Link 2003; Buckley, Metlitski \& Zhitnitsky 2004; but see also
Alford, Good \& Reddy 2005).  

A further possibility is that
unconfined quarks are present and that the long-term stability of
the core magnetic field is then determined by the properties of the Meissner
effect for a colour superconductor (Alford, Berges \& Rajagopal 2000).
If this very interesting question is to be studied by observations on
neutron star magnetic fields it is essential that there are analyses of 
magnetic flux transport in more prosaic systems, such as baryonic
type I and two-gap superconductors.  The present paper is addressed
to this problem.

Magnetic flux transport for type II superconductivity, assuming the matter
composition anticipated at $\rho\simeq\rho_{0}$, has been analysed
previously (see Jones 2006; also references cited therein).  The problem
is well-defined owing to the existence of the mixed state in which, on
microscopic scales, magnetic flux is quantized in units of
$\phi_{0} = hc/2e = 2.07\times 10^{-7}$ G cm$^{2}$ and confined to the
cores of proton vortices.  Type I superconductors behave differently at
the transition from the normal state in the presence of a magnetic field
owing to their positive surface energy, equivalent to the condition
$\kappa < 1/\sqrt{2}$.  The intermediate state formed 
minimizes surface area as much as external constraints allow.  It consists
of flux-free superconductor
in equilibrium with a filamentary structure of normal protons and
magnetic flux density
$B \approx H_{c}$, where $H_{c}$ is the thermodynamic critical field.
Averaged over volumes large compared with the scale of the structure,
the magnetic flux density $\langle {\bf B}\rangle$ is that of the initial
normal system.  Owing to the external constraints, the minimization
condition does not, in itself, lead to structures
that are well-defined or of universal form.  Being of little practical
utility, type I superconductors have been less well studied than type II.
We refer to Tinkham (1996) for further details.

As in the type II case, we assume that an approximation to static
hydrodynamic equilibrium exists in the neutron star interior at times
before the superconducting phase transitions.  This equilibrium changes
because the spatially-averaged components of the type I superconductor
stress tensor are larger than those of the normal-state Maxwell tensor
by a factor of $H_{c}/\langle B\rangle$, which is of the same order as  
for type II superconductors (Jones 1975, Easson \& Pethick
1977).  The change in the stress tensor occurs in a time short compared
with any possible flux expulsion time because cooling of
the neutron star interior is rapid at the superconductor critical
temperature.  The problem is to determine how the magnetic flux
filaments move under any buoyancy force which may then appear.

The presence of $\Sigma^{-}$ hyperons at $\rho\sim 2\rho_{0}$ would
make the neutron star interior an example of the two-gap superconductor
which is of current interest in studies of liquid metallic hydrogen
at high pressures (Babaev, Sudb\o \hspace{0.1cm}\& Ashcroft 2004).
Given that
several baryonic components may be present in the inner core, it 
might be thought that its superconducting properties are complex.
But it appears that the complete system can be a type I superconductor
only if all individual Fermi liquids are either normal or type I.

Both the two-gap case and the factors that determine the form and
scale of the filamentary type I structure are addressed in Section 2.
The various forces acting on a moving type I filament are found in
Section 3.  That arising from the stable stratification of neutron
star matter (see Goldreich \& Reisenegger 1992, who introduced the concept
as a factor limiting non-solenoidal ambipolar diffusion of magnetic flux)
is by far the most
important in all parts of the core where the negative
$\mu$-meson threshold is exceeded. In these regions, filament motion is
limited by the rate of
lepton diffusion between the filament and the flux-free superconductor.
The order of magnitude of the force between a neutron vortex and a type I
filament is evaluated in Section 4. It is shown that neutron vortices
are effectively trapped by the structure, as is the case in type II
superconductivity.
Section 5 gives a summary of the circumstances
under which the inner core could retain magnetic flux for times long
compared with the ages of most pulsars or the spin-down times of the
neutron stars in binary systems.  If these can be excluded, any evidence
of long-term flux freezing would indicate the presence of more exotic
core structures.  In common with the previous paper on type II
superconductivity (Jones 2006) the scope of this paper is intended to be
purely technical and it does not consider observational evidence
relevant to magnetic flux evolution.

\section{Type I and two-gap superconductors}

The intermediate state of a type I superconductor has no universal form
(see Tinkham 1996) and, in a neutron star core, must be largely defined
by the flux distribution in external regions where it can be regarded
as frozen for the short time-scales relevant to the superconducting
transition.  Given that constraint, we anticipate that the
spatially-averaged magnetic flux density $\langle {\bf B}\rangle$
will initially be
identical with the normal-state flux distribution existing before the
transition.  We assume that, in the initial stages of the transition,
small flux-free regions form with a growth rate governed by ohmic
diffusion in the surrounding normal phase.  A cylindrical shape would
minimize surface area and least perturb the surrounding magnetic flux
distribution.  Unless $\langle B\rangle$ is quite close to $H_{c}$,
growth of these flux-free regions must lead to their coalescence and
the formation of a filamentary structure of magnetic flux and normal
protons occupying a fraction $\langle B\rangle/H_{c}$ of spatial
volume.

The temporal order of the various superconductor transitions which
occur is uncertain.  An isothermal neutron star core cannot be assumed
at the early times in which these transitions occur
because its composition may permit direct-Urca processes in some
regions but not others.  However, it is possible to infer that the
filamentary structure is unlikely to be a two-dimensional lattice.
That it is probably interconnected in a complex way can be seen from
the following considerations. Suppose that, at any instant, type I
superconductivity is confined between spherical surfaces of radii
$r_{1,2}(t)$ with $r_{2} > r_{1}$.  As the neutron star cools, these
surfaces move because they are dependent on the evolving temperature
distribution $T(r,t)$ and on the critical temperature $T_{c}(r,t)$.
Filaments with a distribution of cross-sectional area and therefore
varying total flux form between these surfaces. 
The case that $\dot{r}_{1} < 0$ must produce complexity because it
involves the merging of two independently formed filament distributions
subject to flux conservation. (In the simple case of a pure dipole field,
the distributions merging would be those formed on the two sides of the
magnetic equatorial plane.) There would be complete complexity for
$r_{1}(\infty) = 0$ and partial complexity for $r_{1}(\infty) > 0$.
This time-dependence of $r_{1}$ would be consistent with the
baryon-number dependence of calculated proton energy gaps at inner-core
densities (see Pethick \& Ravenhall 1995; Heiselberg \& Hjorth-Jensen
2000). Complex interconnections are also formed as a result of the
fusion of adjacent filaments which move, under the buoyancy force,
with velocities that depend on their radii (see Section 3).
Our assumption about filament orientation is that the spatially-averaged
magnetic flux density $\langle {\bf B}\rangle$ will initially conform
with the pre-transition flux distribution.  Any large-scale rearrangement
at the time of the transition would be energetically not allowed.

The filaments are subject to a post-transition buoyancy force,
${\bf f}_{B}$ per unit volume.  Its origin is more simple than in the
type II case owing to the macroscopic cross-sectional area of the
filaments.  The pre-transition Maxwell tensor expressed in terms of
the spatially-averaged fields contains products
$\langle B_{i}\rangle \langle B_{j}\rangle$.  These are replaced by
$\langle B_{i}B_{j}\rangle$
in which $B\approx H_{c}$.  During motion toward a new
equilibrium, the cross-sectional area of a filament may change owing
to flux conservation and $\rho$-dependence of $\Delta_{p}$ and there
may be inward or outward lepton or proton diffusion, constrained by
electrical neutrality.

If superfluid $\Sigma^{-}$ hyperons are present (the threshold is at
$\rho\sim 2\rho_{0}$ in many equations of state), there is
negligible Josephson coupling with the protons and the system is
analogous with the two-gap superconductor of current interest,
the particular example being that of liquid metallic hydrogen at high
pressure (see Babaev et al 2004).  We can assume that the internal
temperature satisfies $T\ll T_{c}$ throughout most of the time interval
in which the flux distribution evolves.  Thus the complicated phase
diagram of such systems does not concern us and the behaviour is simple.
The reasons for this will be given here in outline only because they
follow very closely the work of Babaev (2002) on the structure of
composite
vortices, to which we refer for further details.  For a system with
several charged Fermi liquids $i=1,2...$, the Ginzburg-Landau
free-energy functional is written down in terms of the condensate
amplitudes $\Psi_{i}$, bare masses $m_{i}$ and charges
$e_{i}$ of the components, and variation with respect to
the vector potential ${\bf A}$ gives the current density ${\bf J}$.
In the approximation that the $\left |\Psi_{i}\right |$ are
position-independent, the condition ${\bf J} = 0$ at radii
$s \gg \lambda$ from a vortex gives a simple expression for
${\bf A}$ in terms of the gradients of the condensate phases
on circular paths about its axis. (The total changes of
phase $\chi_{i}$ on these paths are the phase windings.)
The line integral of ${\bf A}$ gives the enclosed flux
which, for two components with $e_{1} = -e_{2}$ and opposite
phase windings
$\chi_{1} = -\chi_{2}$ of magnitude $\left|\chi_{1,2}\right| = 2\pi$,
is exactly $\phi_{0}$.  This result, a single flux quantum,
holds for any number of components provided $e_{i}^{2}$
and $e_{i}\chi_{i}$ are both constants independent of $i$.
Substitution of ${\bf A}$ into the Ginzburg-Landau functional
shows that, for these phase windings, the long-range kinetic terms vanish
identically and that single flux-quantum composite vortices are the
lowest-lying states of the superconductor that can support magnetic
flux.  (The case of a zero phase winding for one of two components
gives the interesting class of vortices with a fractional quantum of
flux, but substitution of ${\bf A}$ into the free-energy functional
shows that the long-range kinetic terms are then finite and that the vortex
self-energy is much greater than for the single flux-quantum case.
These vortices are not of interest at the temperatures $T\ll T_{c}$
considered here.)  It follows that a system of several components, at
least one of which
satisfies the surface energy condition
$\kappa > 1/\sqrt{2}$, will behave as a type II superconductor even
at magnetic field strengths that exceed the thermodynamic critical
fields of all but that one component (see Babaev 2002).

The conclusion is that if several charged baryonic components are
present, as may be the case in the inner core, the system is a
type II superconductor unless no component satisfies the
$\kappa > 1/\sqrt{2}$ condition.  This simple result indicates that
type I superconductivity may be limited to no more than a small
fraction of the core volume.

\section{The forces on a moving type I filament}

\subsection{Leptonic frictional forces}

Under a buoyancy force ${\bf f}_{B}$, the filamentary structure of a type I
superconductor moves to some extent as a single entity because the nature
of the stress tensor inhibits differential motion and the resulting
filament curvature. To estimate the forces that balance ${\bf f}_{B}$, we
assume that the filaments are locally cylindrical, so minimizing
surface area at constant flux, and are described by cylindrical polar
coordinates ${\bf s}\equiv(s,\theta)$ moving with the filament.
Except where otherwise stated, the radius
$s=a$ is assumed to be macroscopic, large compared with lepton
gyro-radii inside filaments or with scattering mean free paths.
For brevity, we consider baryonic matter composed of protons and
superfluid neutrons only.  Then the relevant mean free path is that
for electron scattering by muons.  We consider, initially, frictional
forces that arise from interaction between filament and leptons.

Easson \& Pethick (1979) have given
the transport relaxation time for electrons charge-neutralized by
normal protons, number densities $N_{e} = N_{p}$.  It is
\begin{eqnarray}
\tau = \frac{12}{\pi^{2}\alpha^{2}}\left(\frac{\epsilon_{Fp}}
{k_{B}T}\right)^{2}\frac{k_{FT}}{ck_{Fe}^{2}}.
\end{eqnarray}
In this expression, $\alpha$ is the fine structure constant;
$k_{Fe}$ and $\epsilon _{Fp}$ are the electron Fermi wave number
and proton Fermi energy.  The screening wave number is determined
by the proton Fermi wave number $k_{Fp}$.  For nonrelativistic
protons, it is,
\begin{eqnarray}
k_{FT} = \left(\frac{4\alpha k_{Fp}m_{p}c}{\pi\hbar}\right)^{1/2}.
\end{eqnarray}
The appropriate adaptation of equation (1) for scattering by
nonrelativistic muons, in the presence of protons, with independent
number densities $N_{e}\neq N_{\mu}\neq N_{p}$ is,
\begin{eqnarray}
\tau_{\mu}^{e} = \frac{12}{\pi^{2}\alpha^{2}}\left(\frac{\epsilon_{F\mu}}
{k_{B}T}\right)^{2}\frac{k_{FT}k_{Fe}^{2}}{ck_{F\mu}^{4}}.
\end{eqnarray}
This is valid for $N_{\mu}$ such that $k_{F\mu}\gg k_{FT}$, and the
optimum definition of $k_{FT}$ is that given by the
most massive particles in the system, the superconducting protons.
This relaxation time is long (of the order of $10^{-12}$ s at
$T=10^{8}$ K), so that the mean free path is always many orders of
magnitude greater than the electron gyro-radius which is
$r_{B} = \epsilon_{Fe}/eB =
3.3 \times 10^{-7} B_{12}^{-1}$ cm, for
$\epsilon_{Fe} = 100$ MeV, where
$B_{12}$ is the magnetic flux density in units of $10^{12}$ G.
The appropriate transport relaxation time $\tau^{e}_{p}$ for electron
scattering by protons in the presence of nonrelativistic muons is
given by an expression identical with equation (3) except that proton
kinematic variables replace those for the muon.

The assumption made here is that the boundary condition satisfied by
the lepton fluid velocity ${\bf v}_{l}$ is
$({\bf v}_{l})_{\perp} = 0$ at $s=a$ on the filament surface.  This differs
from the condition ${\bf v}_{l}(a) = 0$ for conventional viscous flow
which assumes, effectively, that a moving particle completely transfers 
its parallel momentum component to a surface on collision.  This is the
basis, for example, of the treatment of flow along a pipe under the Knudsen
condition of kinetic theory (see Kennard 1938). However, there is negligible
transfer of the parallel component in the case of lepton interaction with
a filament, effectively a cylinder of magnetic flux density
$B\approx H_{c}$.  The small transfer which does occur is a consequence
of scattering by leptons or protons.  But the lepton mean free path is
many orders of magnitude larger than the gyro-radius and becomes infinite
in the $T\rightarrow 0$ limit. On the basis of these considerations,
we shall ignore the moving surface as a source of vorticity, which
it would be in conventional viscous flow, and assume that the lepton
flux $N_{l}{\bf v}_{l}$ is an irrotational and solenoidal vector. Its
potential satisfies Laplace's equation and is determined by a Neumann
boundary condition (see Batchelor 1967) on a static
surface in coordinates fixed in the rotating star but instantaneously
coincident with those of the moving filament.
The velocity ${\bf v}_{l}$ is easily
obtained for the case of constant number density $N_{l}$.  In this frame,
it has the familiar dipole form, with components
$(v_{l})_{s} = -Ua^{2}\cos\theta/s^{2}$ and
$(v_{l})_{\theta} = -Ua^{2}\sin\theta/s^{2}$, for filament velocity
${\bf U}$ perpendicular to its axis.
By integration of the expression for the dissipation
rate per unit volume given in terms of the stress tensor
(see Landau \& Lifshitz 1959), we find that the
electronic viscous force per unit length acting on a filament is,
\begin{eqnarray}
{\bf f}_{e} = -8\pi\eta_{e}{\bf U},
\end{eqnarray}
in which the shear viscosity is,
\begin{eqnarray}
\eta_{e} = \frac{1}{15}\tau^{e}_{\mu}N_{e}\epsilon_{Fe}
\end{eqnarray}
(Easson \& Pethick 1979) and is of the order of $10^{20}$
g cm$^{-1}$ s$^{-1}$ at $10^{8}$ K.
It is worth noting that the same calculation, made for a sphere of
radius $a$, gives a force $-12\pi\eta_{e}a{\bf U}$ which is twice the
Stokes force, an example of the minimum dissipation theorem
(see Batchelor 1967, p. 227).
Equation (4) is independent of $a$
provided $a\gg c\tau^{e}_{\mu}$ and so can be significant for thin
filaments.  For smaller values of $a$, but such that $a\gg r_{B}$,
the appropriate expression for the force can be found by an
elementary classical kinetic theory calculation.  The filament is
regarded simply as a cylinder of magnetic flux density $B\approx H_{c}$.
The electron momentum and flux vectors are transformed into the
filament rest frame
and the momentum transfer obtained directly by integration
over its surface.  Under this condition, the force is
given by,
\begin{eqnarray}
{\bf f}_{e} = - \frac{\pi}{2}N_{e}a\hbar k_{Fe}{\bf U}
\end{eqnarray}
and is temperature-independent.  Forces generated by interaction with
muons are given by expressions of the same form as equations (4)-(6)
subject to replacement of electron number density 
with that for the muons, $k_{Fe}$ by $k_{F\mu}$, and $\epsilon_{Fe}$
by the muon kinetic energy $\epsilon_{F\mu}$.
For nonrelativistic muons, the relaxation time,
\begin{eqnarray}
\tau^{\mu}_{e} = \tau^{e}_{\mu}\left(\frac{N_{\mu}
m_{\mu}c^{2}}
{N_{e}\epsilon_{Fe}}\right),
\end{eqnarray}
replaces that in equation (5).

Equations (4) and (6) have been obtained under the assumption of
constant $N_{l}$ but
it is worth mentioning that a further frictional force arises if the
scale length $\mathcal{L}$ for variation of a leptonic number density,
probably $N_{\mu}$, with depth is
small compared with the neutron star radius.  The
reason is that the distributions of the fluid velocities $v_{e}$ and
$v_{\mu}$ produced by filament movement are then not exactly identical.
Relative motion of the two fluids gives a force per unit length of
filament of the order of,
\begin{eqnarray}
{\bf f}_{e\mu} = - \frac{\pi N_{e}\epsilon_{Fe}a^{4}}
{\tau^{e}_{\mu}c^{2}\mathcal{L}^{2}}{\bf U}.
\end{eqnarray}
It is negligible at small $a$ compared with ${\bf f}_{e}$, and at large
$a$ is less significant than the force arising from the stable
stratification of neutron star matter whose order of magnitude will
be calculated here in much more detail.

\subsection{Stratification and stability}

It is assumed here that, immediately following the superconducting
transition and before filament movement, the conditions
for charge neutrality and weak-interaction equilibrium in terms of
number densities and chemical potentials, $N_{p} = N_{e} + N_{\mu}$
and $\mu_{n} - \mu_{p} = \mu_{e} = \mu_{\mu} + m_{\mu}$, are
satisfied in the filaments and in the flux-free superconductor.
With this definition of muon chemical potential, which excludes rest
mass, the pressure can be expressed as a sum of its components,
\begin{eqnarray}
P = P_{h} + P_{e} + P_{\mu} 
    = P_{h} + \frac{1}{4}N_{e}\mu_{e} + \frac{2}{5}N_{\mu}\mu_{\mu},
\end{eqnarray}
where $P_{h}$ is the baryonic component and $P_{e,\mu}$ are the leptonic
components.  We consider the effect of
a small but finite outward displacement $r\rightarrow r + \delta r$
in the position of
a filament on the equilibrium of the matter in its interior. 
The initial problem is to determine the effect of the movement
on the leptons and superconductor outside the filament.  As
noted in Section 3.1,
the resulting particle fluxes $N_{p,e,\mu}{\bf v}_{p,e,\mu}$, defined by a
Neumann boundary condition at the surface of the moving filament, are
irrotational and solenoidal.  Thus the number densities $N_{p,e,\mu}$
remain unchanged as functions of position in a frame of reference
fixed to the rotating
star even though weak-interaction transition rates are negligibly small.
A displacement over the same distance $\delta r$ within the proton
superconductor outside the filament is associated with number density
changes $N_{n}\rightarrow N_{n}(1-\zeta_{n})$ etc in each particle
type and with a pressure change,
\begin{eqnarray} 
\delta P = - Q_{n}\zeta_{n} - Q_{p}\zeta_{p} - \frac{4}{3}P_{e}\zeta_{e}
- \frac{5}{3}P_{\mu}\zeta_{\mu},
\end{eqnarray}
in which the coefficients are defined as
$Q_{n,p} = N_{n,p}(\partial P/\partial N_{n,p})$.
Our assumption is that the equilibrium conditions are always
satisfied everywhere in the superconductor outside the filaments.
Thus we can eliminate
$\zeta_{p,e,\mu}$ and express $\delta P$ as a linear
function of $\zeta_{n}$,
\begin{eqnarray}
\delta P = - \zeta_{n}\left(Q_{n} + \left(Q_{p}C_{pe}+\frac{4}{3}
P_{e}+\frac{5}{3}P_{\mu}C_{\mu e}\right)C_{en}\right),
\end{eqnarray}
in which the parameters are defined as follows;
\begin{eqnarray}
C_{\mu e} = \frac{\mu_{e}}{2\mu_{\mu}},
\end{eqnarray}
\begin{eqnarray}
C_{pe} = \frac{N_{e} + N_{\mu}C_{\mu e}}{N_{p}},
\end{eqnarray}
\begin{eqnarray}
C_{en} = \frac{A_{n}}{\frac{1}{3}\mu_{e} - A_{p}C_{pe}},
\end{eqnarray}
where,
\begin{eqnarray}
A_{n} = N_{n}\frac{\partial(\mu_{n}-\mu_{p})}{\partial N_{n}},
\end{eqnarray}
and
\begin{eqnarray}
A_{p} = N_{p}\frac{\partial(\mu_{n}-\mu_{p})}{\partial N_{p}}.
\end{eqnarray}
But inside the displaced filament, the constraint of
electrical neutrality, in the absence of diffusion and of significant
weak-interaction transition rates, requires $\zeta_{p} = \zeta_{e}
= \zeta_{\mu} = \tilde{\zeta}$.  (A tilde denotes quantities whose
values inside a filament may differ from external values.  The precise
relation existing between external and internal values at $r$ would be
replicated, under weak-interaction equilibrium, at $r + \delta r$ and
so does not enter the problem. The effect of variation in $H_{c}$ over
the displacement length $\delta r$ is too small to be significant.)
The further constraints of pressure and neutron equilibrium
are expressed as $\delta P = \delta \tilde{P}$ and
$\delta \mu_{n} = \delta \tilde{\mu}_{n}$, respectively.  Because
$\mu_{n}$ is a function of both $N_{n}$ and $N_{p}$ in general, it is
necessary to define a distinct internal value $\tilde{\zeta}_{n} \neq
\zeta_{n}$ to represent the change in internal neutron number density.

The pressure equilibrium constraint is then,
\begin{eqnarray}
\delta P = \delta\tilde{P} = - \tilde{\zeta}_{n}Q_{n} -
\tilde{\zeta}\left(Q_{p}
+ \frac{4}{3}P_{e} + \frac{5}{3}P_{\mu}\right),
\end{eqnarray}
and the chemical potential constraint is,
\begin{eqnarray}
\left(\frac{\partial\mu_{n}}{\partial N_{n}}\right)N_{n}
\left(\zeta_{n} - \tilde{\zeta}_{n}\right) +
\left(\frac{\partial\mu_{n}}{\partial N_{p}}\right)N_{p}
\left(\zeta_{p} - \tilde{\zeta}\right) = 0.
\end{eqnarray}
Equations (11), (17) and (18) can be solved for $\tilde{\zeta} =
G_{c}\zeta_{n}$ and $\tilde{\zeta}_{n} = G_{n}\zeta_{n}$.
The differences between the chemical potentials inside the filament
and those in the flux-free superconductor at the same radius $r+\delta r$
from the centre of the star are then,
\begin{eqnarray}
\delta\mu_{e} = \frac{1}{3}\mu_{e}\left(G_{c}-C_{en}\right)\zeta_{n},
\end{eqnarray}
\begin{eqnarray}
\delta\mu_{\mu} = \frac{2}{3}\mu_{\mu}\left(G_{c}-C_{\mu e}C_{en}\right)
\zeta_{n},
\end{eqnarray}
\begin{eqnarray}
\delta\mu_{p} & = & \left(\frac{\partial\mu_{p}}{\partial N_{n}}
\right)N_{n}\left(G_{n}-1\right)\zeta_{n}   \nonumber \\ 
			  &   & + \left(\frac{\partial\mu_{p}}{\partial N_{p}}
\right)N_{p}\left(G_{c}-C_{pe}C_{en}\right)\zeta_{n}.
\end{eqnarray}
The pressure increment $\delta P$ given by equation (11) can also, of
course, be expressed simply in terms of the local matter density and
gravitational acceleration $g$,
\begin{eqnarray}
\delta P = g\left(N_{n} + N_{p}\right)m_{n}\delta r,
\end{eqnarray}
so that, from equations (11) and (19)-(22), the energy excess per unit
volume of matter inside a filament can be expressed in terms of a force
constant $K$,
\begin{eqnarray}
\frac{1}{2}\sum_{i}\left(\frac{\partial N_{i}}{\partial \mu_{i}}\right)
\left(\delta \mu_{i}\right)^{2} =
\frac{1}{2}K\left(\delta r\right)^{2}.
\end{eqnarray}
This equation defines $K$ in terms of the composition-dependent parameters
contained in (11) and (19)-(22), but the expression so obtained is not
compact.  Its evaluation is clearly very
dependent on the choice of equation of state.  The order of magnitude,
for non-interacting Fermi gases (chemical potentials $\mu_{n} = 128.2,
\mu_{p} = 8.2, \mu_{e} = 120, \mu_{\mu} = 14.3$ MeV) gives
$K = 6\times 10^{20}g_{14}^{2}$ erg cm$^{-5}$.  This is an interesting
result.  It shows that filament movement under a buoyancy force $f_{B}$
would be stopped in a distance $\delta r \approx f_{B}/K\sim 1$ cm.

This metastable state formed inside the filament can, in principle,
convert to the equilibrium state by the irreversible loss of neutrino
and thermal energy.  But for superfluid and superconducting systems,
baryonic semi-leptonic transition rates are negligibly small at
$T \sim 10^{8}$ K, unlike the normal Fermi systems in the paper of
Goldreich \& Reisenegger, and they will not be considered further here.
Direct leptonic transitions $\mu \rightleftharpoons e$ with neutrino
emission are excluded by the large difference between electron and
muon Fermi momenta.
The most significant process affecting the interior of a filament
appears to be the diffusion of leptons and protons to or from the
proton superconductor.  These rates determine the filament velocity.

\subsection{Diffusion and filament velocity}

This diffusion, which was neglected in Section 3.2, means that the
chemical potential differences given by equations (19)-(21) are actually
functions of position, $\delta \mu_{i}(s)$ for $0 < s < a$.
The boundary condition $\delta \mu_{i}(a) = 0$
is a consequence, for protons, of the strong interaction and, for
leptons, of the fact that mean free paths in the flux-free
superconductor at $s > a$ are always many orders of magnitude larger
than their gyro-radii at $s < a$.  We are
unaware of any full calculations of this diffusion process valid for
$B \approx H_{c}$ and so have been obliged to rely on an order of
magnitude estimate.  We begin by noting that the field is
marginally non-quantizing, as defined by Potekhin (1999), at
$T = 10^{8}$ K and $B = 10^{14}$ G.  A lepton typically completes
many classical revolutions between scatters and the gyro-radius is
many orders of magnitude greater than the Fermi wavelength.
It then follows that the total electron and muon collision rates per
unit volume are independent
of $B$ and are given by equation (3) and by an equation of the same
form in which the muon kinematic variables are replaced by those for
the protons.  We shall consider the electron case and assume that
a scatter causes a classical guiding centre displacement of the
order of $r_{B}k_{FT}/k_{Fe}$, where $r_{B}$ is the gyro-radius.
The radial electron flux is of the order of,
\begin{eqnarray}
J^{e} & = & - N_{e}\left(\frac{1}{\tau^{e}_{p}} +
 \frac{1}{\tau^{e}_{\mu}}\right)
\left(\frac{r_{B}k_{FT}}{k_{Fe}k_{B}T}\frac{\partial \delta \mu_{e}}
{\partial s}\right)\left(\frac{r_{B}k_{FT}}{k_{Fe}}\right) \\ \nonumber
      & = & - D\frac{\partial N_{e}}{\partial s},
\end{eqnarray}
and defines the diffusion coefficient $D$,
\begin{eqnarray}
D = \frac{\mu_{e}}{3k_{B}T}\left(\frac{1}{\tau^{e}_{p}} +
\frac{1}{\tau^{e}_{\mu}}\right)\left(\frac{r_{B}k_{FT}}{k_{Fe}}\right)^{2},
\end{eqnarray}
which is a linear function of $T$.  Diffusion is, of course, constrained
by the need to maintain electrical neutrality inside the filament.  To
allow for this,  
equation (24) should be modified to include the radial electric field
necessary to give identical fluxes for leptons and protons, but we have
neglected this problem and use the unconstrained electron diffusion rate
given by equation (24)
for the order of magnitude velocity estimate made here.

The equilibrium considered at the end of Section 3.2 continues to exist
in the presence of diffusion because the chemical potential differences
$\delta \mu_{i}$ are maintained by outward movement of the filament.
For a cylinder of radius $s < a$ inside the filament, the continuity
equation is,
\begin{eqnarray}
\int^{s}_{0}\frac{\partial\delta \mu_{e}}{\partial t}2\pi s^{\prime}
ds^{\prime} = 2\pi s D\left(\frac{\partial \delta \mu_{e}}
{\partial s}\right)_{s} + \pi s^{2}S,
\end{eqnarray}
in which the source term $S$ depends on filament velocity $U$,
\begin{eqnarray}
S \approx \frac{UK}{f_{B}}\delta \mu_{e}(0).
\end{eqnarray}
The condition for a time-independent $\delta \mu_{e}$ is,
\begin{eqnarray}
\frac{\partial \delta \mu_{e}}{\partial s} = - \frac{sS}{2D},
\end{eqnarray}
which, with the boundary condition $\delta \mu_{e}(a) = 0$, defines
a steady-state velocity,
\begin{eqnarray}
U = \frac{4Df_{B}}{Ka^{2}}.
\end{eqnarray}
Evaluation for the parameters assumed in Section 3.2 gives a diffusion
coefficient $D = 2.1\times 10^{-1}T_{8}\left(H_{c14}\right)^{-2}$
cm$^{2}$ s$^{-1}$, where $T_{8}$ is the temperature in units of
$10^{8}$ K and $H_{c14}$ is the thermodynamic critical field in
units of $10^{14}$ G.  The velocity is,
\begin{eqnarray}
U = \frac{0.14}{a^{2}}g_{14}^{-2}T_{8}f_{B20}\left(H_{c14}\right)^{-2},
\end{eqnarray}
in units of cm s$^{-1}$.
It is of a significant magnitude for filamentary structures with
size less than $a \sim 10^{4}$ cm.

\subsection{The Magnus force}

All the forces considered above are frictional in character, and it
remains to note the presence of a Magnus force on the moving filament.
The proton superfluid circulates around a stationary filament of radius
$a$ at a velocity,
\begin{eqnarray}
{\bf v}_{p} = - \frac{e}{m_{p}c}{\bf A}.
\end{eqnarray}
We consider the case $a \gg \lambda$, where $\lambda$ is the proton
penetration depth, for which the vector potential can be chosen as,
\begin{eqnarray}
A_{\theta} = -H_{c}\lambda \exp\left(-\frac{s-a}{\lambda}\right).
\end{eqnarray}
For a filament moving with velocity ${\bf U}$, the circulating superfluid
velocity is changed by the addition of a small increment
$\delta {\bf v}_{p}$.  Flow is irrotational and solenoidal and is
subject to the boundary condition $v_{p\perp} = 0$
at $s=a$, so that the increment is $\delta v_{p\theta} = 2U\sin\theta$
at $s=a$ in the filament rest frame.  Neglecting entrainment terms,
which do not contribute
here, the isotropic component of the superfluid stress tensor,
\begin{eqnarray}
T^{p}_{ij} = \rho_{s}^{pp}v_{pi}v_{pj}
 - \frac{1}{2}\rho_{s}^{pp}v_{p}^{2}\delta_{ij},
\end{eqnarray}
for superfluid density $\rho_{s}^{pp}$,
is changed by a small increment
$- \rho_{s}^{pp}v_{p\theta}\delta v_{p\theta}$.
Integration of this over the filament surface gives a transverse force
of magnitude,
\begin{eqnarray}
f_{M} = \rho_{s}^{pp}U\left(\frac{2\pi aeH_{c}\lambda}{m_{p}c}\right),
\end{eqnarray}
per unit length of filament in which, as
expected, the final bracketed term is the circulation.

This force is not large compared with $f_{B}$ for macroscopic values of
$a$ and in any case, is cancelled by the lepton flow which is subject to
the same boundary condition and, as we have argued in Section 3.1, is
almost exactly irrotational.  That it is not
exactly irrotational and our neglect of quasi-particle excitations within
the filament both imply that this cancellation cannot be exact.  But these
problems are not of significance for the results obtained in this paper.

\section{Interaction between neutron vortices and type I filaments}

It is generally assumed that, in regions of type II proton superconductivity
and neutron superfluidity, the movement of magnetic flux is constrained by
the strong and electromagnetic interaction between neutron and proton
vortices (Sauls 1989).  Flux movement is also highly constrained in type I
regions.  This can be seen by calculating the change in free energy per
unit length arising from the movement of a neutron vortex from the flux-free
region of a type I proton superconductor to a position co-axial with
a normal proton filament of magnetic flux density $B\approx H_{c}$.  We
shall assume that the filament radius $a$ is at least of the order of
the neutron intervortex spacing ($10^{-3}-10^{-2}$ cm).  The free energy
density for a system of neutron and proton superfluids can be expressed
in the Ginzburg-Landau form as a functional of the condensate
amplitudes $\Psi_{n,p}$.  It includes a term representing the phenomenon 
of superfluid entrainment, originally introduced by Andreev \& Bashkin (1976)
in connection with solutions of He$^{3}$ in He$^{4}$, and further extended
to the magnetohydrodynamics of neutron-star superfluids by Vardanian \&
Sedrakian (1981).  It is likely that the inner-core neutrons form a
$^{3}P_{2}$ superfluid, but for the order of magnitude evaluation made here
we shall neglect this complication and assume $^{1}S_{0}$ structure.
In the approximation that
the $\left|\Psi_{n,p}\right|$ are position-independent, it reduces to the
intuitive form,
\begin{eqnarray}
f_{GL} = f_{c}^{p} + f_{c}^{n} + \frac{1}{2}\rho^{pp}_{s}{\bf v}_{p}^{2} + 
\rho^{pn}_{s}{\bf v}_{p}\cdot{\bf v}_{n}
 + \frac{1}{2}\rho_{s}^{nn}{\bf v}_{n}^{2}  \nonumber  \\
 + \frac{1}{8\pi}\left(\nabla \times {\bf A}\right)^{2}
 \end{eqnarray}
(see, for example, Alpar, Langer \& Sauls 1984), in which $f_{c}^{p,n}$ are the
proton and neutron condensation energy densities.  The superfluid velocities are,
\begin{eqnarray}
 {\bf v}_{n} &=& \frac{\hbar}{2m_{n}}\nabla \chi_{n}  \\  \nonumber
 {\bf v}_{p} &=& \frac{\hbar}{2m_{p}}\nabla \chi_{p} - \frac{e}{m_{p}c}{\bf A},
\end{eqnarray}
where $\chi_{n,p}$ are the condensate phases and $m_{n,p}$ the bare masses.
We refer to Alpar et al for the definitions of the superfluid densities
$\rho_{s}^{nn,pp}$ and of the entrainment term $\rho_{s}^{pn}$ in terms of
the bare and effective particle masses.
Variation with respect to ${\bf A}$ gives the electromagnetic current density,
\begin{eqnarray}
 {\bf J} = \frac{e}{m_{p}}\left(\rho_{s}^{pp}{\bf v}_{p} + \rho_{s}^{pn}
 {\bf v}_{n}\right).
\end{eqnarray}
It will be convenient here to re-use the cylindrical polar coordinates
$(s,\theta)$ to describe neutron vortex structure.  The condition
${\bf J}=0$ at $s\gg \lambda$ fixes the proton superfluid velocity
${\bf v}_{p}$ and, for zero proton winding phase, the asymptotic vector
potential,
\begin{eqnarray}
 A_{\theta} = \frac{\rho_{s}^{pn}}{\rho_{s}^{pp}}\frac{\hbar c}{2e}
 \frac{m_{p}}{m_{n}}\frac{1}{s} = A_{0}\lambda/s
\end{eqnarray}
whose line integral gives the fractional quantum of magnetic flux $\phi$
described by Alpar et al.
 
Movement of the neutron vortex from the flux-free region of superconductor
transfers its fractional flux quantum to the filament.  There is a filament
volume change which depends on the relative orientation of the vortex flux
and the magnetic flux ${\bf B}$ in the filament.  This depends both on the
spin direction of the star and on the sign of the entrainment
density $\rho_{s}^{pn}$, and so is unknown.
To be specific, we shall assume here, initially, the parallel case.
Because the condensation
energy density and $H_{c}$ are related by $-f_{c}^{p} = H_{c}^{2}/8\pi$, the
filament free energy is increased by $\phi H_{c}/4\pi$ per unit length.
There is no proton supercurrent inside the filament. Thus
the total increase in free energy per unit length is the sum of this term
and the change in the volume integral of $f_{GL}$,
\begin{eqnarray}
\Delta E = \frac{\phi H_{c}}{4\pi} - \hspace{5cm} \nonumber  \\
  \hspace{0.5cm}        \int^{b}_{0}2\pi s\,ds\left(\frac{1}{2}
\rho_{s}^{pp}v_{p}^{2} + \rho_{s}^{pn}{\bf v}_{p}\cdot{\bf v}_{n} +
\frac{1}{8\pi}\left(\nabla \times {\bf A}\right)^{2}\right)
\end{eqnarray}
in which the upper limit of integration is defined by the spacing between
neutron vortices.  Its order of magnitude can be evaluated for a model
vortex defined by,
\begin{eqnarray}
A_{\theta} & = & A_{0}s/\lambda,  \hspace{2cm}   0<s<\lambda,  \nonumber  \\
A_{\theta} & = & A_{0}\lambda/s,  \hspace{2cm}     b>s>\lambda, \nonumber  \\
v_{n} & = & v_{n}^{0}s/\xi_{n},   \hspace{1.9cm}   0<s<\xi_{n}, \nonumber  \\
v_{n} & = & v_{n}^{0}\xi_{n}/s,   \hspace{1.9cm}     b>s>\xi_{n},
\end{eqnarray}
where $\xi_{n}$ is the neutron coherence length and
$v_{n}^{0} = \hbar/(2m_{n}\xi_{n})$.  In this model, the vortex has
flux $\phi = |y|\phi_{0}$, where $y = \rho_{s}^{pn}/\rho_{s}^{pp}$.
Entrainment reduces vortex self-energy independently of the sign of
$y$.  For typical values ($H_{c} = 10^{15}$
G, $\ln\,b/\lambda =10$, $\xi_{n} \ll \lambda$) and the proton number density
assumed in Section 3.2, the energy difference per unit length of vortex is
$\Delta E = 1.6\times 10^{7}|y| + 4.6\times 10^{7}y^{2}$ erg cm$^{-1}$.
The first term would become negative if
the vortex flux and ${\bf B}$ were antiparallel.  The parameter $y$ is
probably smaller than unity but is
not well known, particularly at the high baryon densities of the inner
core, so that $\Delta E$ could, in principle,  be of either sign.
But its magnitude is large, of the order of $10^{7}$ erg cm$^{-1}$
or $0.6$ MeV fm$^{-1}$.  At this number density, the penetration depth is
$\lambda = 8\times 10^{-12}$ cm and the interaction force,
$\Delta E/\lambda \sim 10^{18}$ dyne cm$^{-1}$, is some orders of magnitude
larger than the Magnus force $\sim 10^{12}\langle v_{n}\rangle$
dyne cm$^{-1}$ arising from
any plausible value of the spatially-averaged neutron superfluid velocity
$\langle v_{n}\rangle$ relative to pinned vortices.  Except for the possibility
of sliding motion, flux movement is constrained in type I regions, as in
type II.

The interaction energy estimated here leads to conclusions differing from
those of Sedrakian,\, Sedrakian \& Zharkov (1997) who have previously
considered the equilibrium state of neutron vortices in a proton type I
superconductor with finite magnetic flux, parallel with the angular
velocity vector of the star.  These authors assumed that the
vortices are surrounded by coaxial normal filaments of radius $a$ and
uniform magnetic flux density, with flux-free superconductor at radii
$a < s < b$, where $b$ is determined by the intervortex spacing.
They obtained $a$ as a function of $b$ and $H_{c}$.
However, entrainment reduces vortex self-energy so that
the case $\Delta E > 0$ is the more probable.  The stable equilibrium is
then a state in which the vortices are surrounded by flux-free
(except for the vortex core) superconductor at $0 < s < \tilde{a}$
and the normal protons and magnetic flux are confined between coaxial
cylindrical surfaces, $\tilde{a} < s < b$.

\section{Conclusions}

In Section 2 we have observed, following recent work on two-gap
superconductors (Babaev 2002; Babaev, Sudb\o\, \& Ashcroft 2004), that a
type I distribution of magnetic flux in a system of several charged
baryonic components is possible only if no superconducting component
satisfies the $\kappa > 1/\sqrt{2}$ condition. Type I characteristics
are perhaps less probable than the ordered vortex-lattice type II
structure for this reason. Nevertheless, it is possible that type I
regions are present and the purpose of this paper is to see how magnetic
flux distributions evolve within them.

Our assumptions about the nature of the superconducting transition have
been stated in Section 2.  A filamentary structure of magnetic flux and
normal baryons forms occupying a fraction $\langle B\rangle/H_{c}$ of
spatial volume. For a small window of possible $\langle B\rangle$ values,
this fraction so large that the structure is inverted, the filaments being
flux-free.  In this case, the magnetic flux evolution would be that of a
normal Fermi system and we refer to Goldreich \& Reisenegger (1992) for
discussions of ambipolar diffusion and of the Hall effect. In particular,
the force derived in Section 3.2 is present above the muon threshold and
removes the possibility of ambipolar diffusion because the lepton
diffusion considered in Section 3.3 is very slow over distances of the
order of the neutron star radius.

The movement of magnetic flux in a type I region is not, of course,
independent of its behaviour elsewhere in the star (and vice-versa).
This inconvenient mutual dependence
is simply the result of flux conservation and of the large increase
in energy that would be the consequence of independent movement in
different regions.  The problem is therefore extremely untidy:
estimates of the filament velocity ${\bf U}$ under the buoyancy force
${\bf f}_{B}$ are made under the unknown constraints imposed by
external regions.

We consider the case in which $\langle B\rangle/H_{c}$ is perhaps an
order of magnitude smaller than unity.  
If the filament size distribution has very small values of $a$, it might
be thought that the system should resemble an irregular type II
vortex lattice with varying (large) integral numbers of flux quanta.
We refer to Jones (2006) and references therein for further
details of magnetic flux movement in a type II superconductor.
But there are some essential differences. It is not a lattice
owing to the complex interconnections described in Section 2, and
filament fusion is favoured because it reduces surface energy.
The significant forces acting at small $a$ are those given by equations
(4) and (6).  Here, rapid diffusion of leptons and protons reduces
chemical potential differences to negligible values so that stable
stratification has no effect.  The buoyancy force per
unit length of filament is $\pi a^{2}{\bf f}_{B}$ and it follows that
the values of $U$  are $a$-dependent.  These
differential values of $U$ lead to contact, fusion of filaments
to larger $a$, and increased values of $U$.  But eventually, larger
$a$ and longer diffusion times cause stable
stratification to become important.
At an internal temperature $T \approx 10^{9}$ K appropriate for a
young neutron star, the velocity $U$ is determined by equation (30)
for filament sizes larger than $a\sim 1$ cm.  Therafter, the progress
of fusion continues throughout the type I region
until it is stopped either by the external factors we have discussed
above or by filament velocities becoming too small.

External factors are the more important constraint limiting the size
of $a$.  For macroscopic values, the decrease in surface energy
through fusion is quite negligible compared with the increase in energy
arising from any large-scale curvature or distortion of an individual
filament. For example, under the buoyancy force ${\bf f}_{B}$ alone,
the equilibrium
central displacement of a filament of length $L$, fixed at its ends by
external constraint, would be limited and of the order of
$8\pi L^{2}f_{B}/H_{c}^{2}$.  This means that if neutron stars have type I
and II regions in concentric radial shells, flux movements within them
are not independent. 

Type I filamentary structures interact strongly with
neutron vortices, most of the interaction energy $\Delta E$ arising from
superfluid entrainment.  For small $a$, the structures move easily
under the buoyancy force ${\bf f}_{B}$ or a force ${\bf f}_{V}$ derived
from interaction with neutron vortices, as do proton vortices in the
type II case.  Reference to equation (30) shows that comovement with
vortices at radial velocities of $10^{-7} - 10^{-6}$ cm s$^{-1}$
occurs easily.  Velocities of these orders of magnitude are possible
during the spin-down of young isolated pulsars, such as the Crab, or
in the propeller-phase spin-down of neutron stars in binary systems.
It is, however, unfortunate that our conclusion depends crucially on
the value of $a$ and on the diffusion coefficient $D$, of which
equation (25) is no more than an order of magnitude estimate.
For values larger than $a\sim 10^{3} - 10^{4}$ cm, the frictional
force produced by diffusion becomes large enough to limit these
velocities. As $a$ approaches its limiting value
($\sim 5\times 10^{5}$ cm) the speed of flux movement tends to that
expected in a normal proton system as described by Goldreich
\& Reisenegger (1992) and referred to earlier in this Section.

\section*{Acknowledgments}

The author thanks Dr. Egor Babaev for a helpful comment on
fractional-flux vortices.

\bsp

\label{lastpage}

\end{document}